\documentclass{article}
\usepackage{spconf,amsmath,graphicx}
\usepackage{amsfonts}
\usepackage{booktabs}
\usepackage{subfigure}
\usepackage{threeparttable}
\usepackage{amsmath}
\title{Binaural Rendering of Ambisonic Signals by Neural Networks}
\name{Yin Zhu\textsuperscript{1*}\footnotemark[1], 
Qiuqiang Kong\textsuperscript{2}, 
Junjie Shi\textsuperscript{2}, 
Shilei Liu\textsuperscript{2}, 
Xuzhou Ye\textsuperscript{2}, 
Ju-chiang Wang\textsuperscript{2}, 
Junping Zhang\textsuperscript{1}}
\address{\textsuperscript{1}Department of Computer Science, Fudan University\\
\textsuperscript{2}Audio \& Music Intelligence (SAMI), ByteDance Inc. \\
\textsuperscript{1}yinzhu20@fudan.edu.cn
}
\begin{document}
%
\maketitle
\renewcommand{\thefootnote}{\fnsymbol{footnote}}
\footnotetext[1]{Work done during internship at ByteDance.}
\begin{abstract}
Binaural rendering of ambisonic signals is of broad interest to virtual reality and immersive media. Conventional methods often require manually measured Head-Related Transfer Functions (HRTFs). To address this issue, we collect a paired ambisonic-binaural dataset and propose a deep learning framework in an end-to-end manner. Experimental results show that neural networks outperform the conventional method in objective metrics and achieve comparable subjective metrics. To validate the proposed framework, we experimentally explore different settings of the input features, model structures, output features, and loss functions. Our proposed system achieves an SDR of 7.32 and MOSs of 3.83, 3.58, 3.87, 3.58 in quality, timbre, localization, and immersion dimensions.
\end{abstract}
\section{Introduction}
\label{sec:intro}

Binaural rendering of Ambisonic signals \cite{moller1992fundamentals} is a task to render binaural signals from ambisonic signals \cite{gerzon1973periphony}. Binaural rendering aims to provide users with headphones a feeling of immersive acoustic, and ambisonic is a surround sound format that has been widely used for binaural rendering in recent years. What distinguishes ambisonic from other surround sound formats is that the channels of ambisonic are spherical harmonics coefficients instead of microphone-recorded signals. 

Head-Related Transfer Function (HRTF) is a response that characterizes how an ear receives a sound from a point in space and is widely used for binaural rendering. The challenge of ambisonic-based binaural rendering is that ambisonic is not in the scope of HRTF. One way is to transform ambisonic to the scope of HRTF, which is called the virtual loudspeaker method \cite{begault20003, mckeag1996sound,  puomio2017optimization}. The basic idea is to assume that there are multiple virtual speakers in the space, and then calculate the signals carried by the virtual speakers separately according to ambisonic. The other is to transform HRTF to the scope of ambisonic, which is called the spherical HRTF (sp-HRTF) method \cite{kirkeby1999digital, ben2021binaural, zaunschirm2018binaural, schorkhuber2018binaural, engel2021improving}. The basic idea is to perform Spherical Fourier Transform (SFT) on HRTFs to get sp-HRTFs. Then, the sp-HRTFs are matched with ambisonic signals to output binaural signals.

However, there are still issues with conventional methods. Firstly, both types of conventional methods require a costly measured HRTF dataset. Secondly, the virtual loudspeaker method amplifies the flaws of ambisonic. Theoretically, an ambisonic of infinite order is a lossless representation of the sound field, but the actual order of ambisonic is finite. Thirdly, the sp-HRTF method suffers from the ``spatial leakage" effect stemmed from SFT~\cite{hold2019improving} and distortion stemmed from order reduction~\cite{kirkeby1999digital, ben2021binaural}. To compensate the distortion, some works based on the Duplex theory \cite{strutt1907our} are proposed \cite{zaunschirm2018binaural, schorkhuber2018binaural, engel2021improving}. Recently, many neural network-based methods demonstrated the ability to process spatial audio. Richard \cite{richard2021binaural} also used TCN to synthesize binaural speeches from mono audio. Leng \cite{leng2022binauralgrad} further used Diffusion Probabilistic Model to synthesize. Nevertheless, those works mainly focus on synthesizing binaural speeches from mono audios. 
There is a lack of research on binaural rendering from ambisonic signals using neural networks.


In this work, we thus propose an end-to-end framework to address the ambisonic-based binaural rendering problem. First, we collect and release a 1-hour Ambisonic and binaural signals dataset simultaneously recorded in a music rehearsal room. Second, we propose to use both the complex spectrum and the phase difference of ambisonic as the input feature. Third, we propose to utilize a variety of neural network architectures, including dense neural networks, recurrent neural networks, and UNet-based convolutional neural networks to map the input feature to an indirect representation of the spectrogram. Our proposed framework has the following advantages: 
1) Our proposed method no longer requires measuring costly HRTFs at different angles in an anechoic room. 
2) Our proposed method can compensate for the information loss of ambisonic with finite order by supplementing with paired binaural signals. 3) Our proposed systems achieve better signal-to-distortion ratio (SDR) and log spectral distance (LSD), and achieve comparable mean opinion score (MOS) compared to conventional methods.

This paper is organized as follows. Section 2 introduces preliminary knowledge. Section 3 presents our implementation of the proposed framework. Section 4 describes the experimental design and results. Section 5 concludes this work.

\section{Preliminary}
\subsection{Terminologies}
Assume there is a sound field $S(\omega, \Omega)$, where $\omega$ is frequency, and $\Omega \equiv (\theta, \phi) \in \mathbb{S}^2$, $\theta \in (-\frac{\pi}{2}, \frac{\pi}{2})$ is elevation angle and $\phi \in [0, 2\pi)$ is azimuth angle. Head-Related Transferred Function (HRTF) can be simply considered as a function of frequency and orientation
$H(\omega, \Omega) = [H^l(\omega, \Omega), H^r(\omega, \Omega)]^T$, indicating left and right ears respectively. Then, the binaural $X(\omega)$ can be obtained by:
\begin{equation}
\label{vanilla ear signals}
    X(\omega) = [X^l(\omega), X^r(\omega)]^T = \int_{\Omega \in \mathbb{S}^2} S(\omega, \Omega) H(\omega, \Omega) d\Omega
\end{equation}
Note that Equation~\eqref{vanilla ear signals} is defined in the frequency domain, and Head-Related Impulse Response (HRIR) is the Fourier transform of HRTF. 
Ambisonic \cite{gerzon1973periphony} is a surround sound format to record sound field. Formally, ambisonic coefficients $A(\omega)$ of order $N_A$ is given by a $(N_A + 1)^2 $ vector, and the $i$-th component of $A(\omega)$ is defined as:
\begin{equation}
    A(\omega)_i = \int_{\Omega \in \mathbb{S}^2} S(\omega, \Omega) y_{i}(\Omega) d\Omega
\end{equation}
where $y_{i}(\Omega)$ is the real-valued spherical harmonics \cite{williams1999fourier}.

\begin{figure}[t]
\centering
\includegraphics[width=\columnwidth]{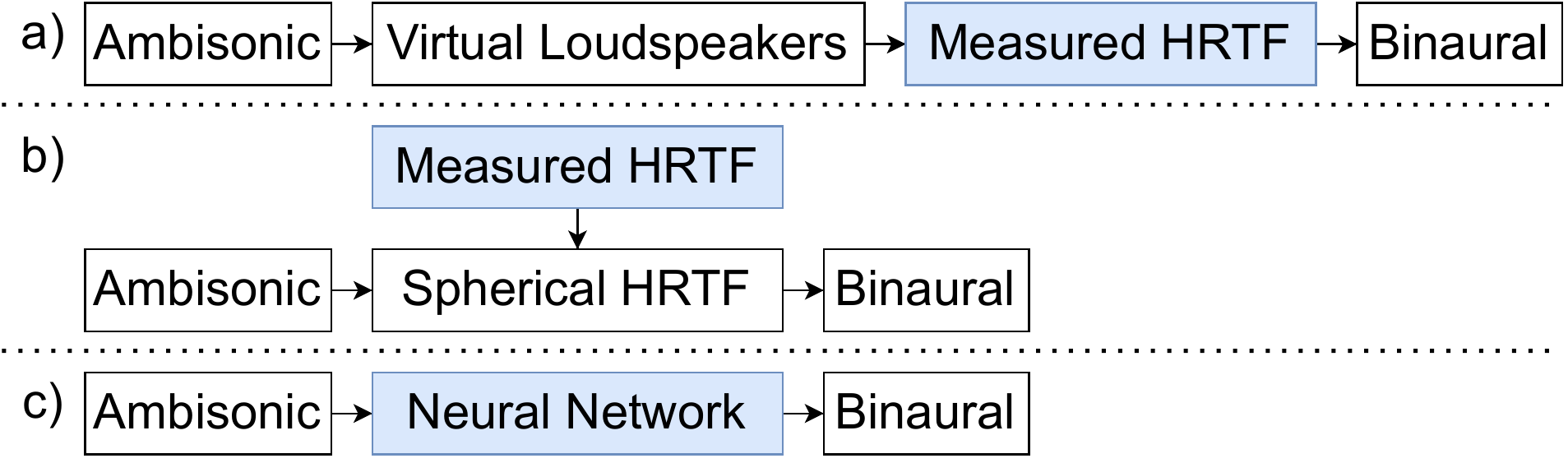}
\caption{
(a) Virtual loudspeaker binaural rendering. (b) sp-HRTF binaural rendering. (c) Neural network binaural rendering.
}
\label{fig:frame}
\end{figure}
\subsection{Formalization}
We will formalize the virtual loudspeaker method and the sp-HRTF method. Fig. \ref{fig:frame}(a) shows the workflow of the virtual loudspeaker method. The virtual loudspeaker method defines a set of decoding functions $\{f_i\}_{i=1, ...N}$ that decodes ambisonic coefficients $A(\omega)$ to the signal of the virtual loudspeaker, where $N$ is the number of virtual loudspeakers. Then the binaural signal $X(\omega)$ is calculated as:
\begin{equation}
    X(\omega) = \sum_{i=1}^{N} f_i(A(\omega)) H_i(\omega)
\end{equation}
where $H_i(\omega) \equiv H(\omega, \Omega_i)$ and $\Omega_i$ depends on the relative angle of $i$-the virtual loudspeaker. Fig. \ref{fig:frame}(b) shows the workflow of the sp-HRTF method. The sp-HRTF method defines a set of encode functions $\{g_i\}_{i=1, ..., (N_A + 1)^2}$ that encodes all HRTFs $\{h(\omega, \Omega)\}_{\Omega \in \mathbb{S}^2}$ to the sphere, and the binaural signal $X(\omega)$ is calculated by:
\begin{equation}
    X(\omega) = \sum_{i=1}^{(N_A+1)^2} A(\omega)_i g_i(\{h(\omega, \Omega)\}_{\Omega \in \mathbb{S}^2})
\end{equation}
\section{Neural Network-based Binaural Rendering}
\label{sec: proposed}
\subsection{Input Feature}
\label{sec:input feature}
We denote the input ambisonic coefficients as $x\in \mathbb{R}^{C \times L}$, where $C$, $L$ denote the channels number and the sequence length of the ambisonic signal $x$, respectively. Our input feature includes a frequency feature and a phase difference feature. First, the human ear is physiologically structured to process different frequencies of sound \cite{beament2001we}. We perform a Short-Time-Fourier-Transformation (STFT) on $x$ to obtain the complex spectrogram $X \in \mathbb{C}^{C \times T \times F}$, where $T$ and $F$ denote the number of time frames and frequency bins respectively. Second, the Duplex theory \cite{strutt1907our} states that lateralization of sound sources is due to the interaural time difference for low frequencies and the interaural level difference for higher frequencies. We calculate the phase differences $\Delta \Phi \in [0, 2\pi)^{\binom{C}{2} \times T \times F}$ between the two channels of $x$, where $\binom{C}{2}$ is the combination of 2 different channels from all channels. Since the radian is not aligned with the complex spectrogram, we further define the phase difference feature as:
\begin{equation}
    D = (||O_B|| \odot \text{cos} \Delta \Phi, ||O_B|| \odot \text{sin} \Delta \Phi)
\end{equation}
where $O_B \in \mathbb{C}^{T \times F}$ is the complex spectrogram of the omnidirectional signal which corresponds to the 0-th order ambisonic coefficient.
\subsection{Output Feature}
\label{sec:output feature}
Experimentally, directly predicting the complex spectrogram will increase the training difficulty. One reason is that small shifts in the phase of signal can make a big difference in the complex spectrogram. Inspired by \cite{kong2021decoupling}, we decouple the estimated magnitude and phase of the complex spectrogram of the binaural signal. Our model outputs the mask $M \in [0, 1]^{B \times T \times F}$ and the phase difference $\sin\angle |M| \in [-1, 1]^{B \times T \times F}$ and $\cos\angle M \in [-1, 1]^{B \times T \times F}$, respectively, where $B$ is the number of output channels and equals 2 for binaural signals.
The magnitude and phase of $ M $ is applied to the complex spectrum of the omnidirectional ambisonic coefficient $ O_B $:
\begin{equation}
    \begin{aligned}
        & X = \mathcal{F}(x) \\
        & |M|, \text{cos} \angle M, \text{sin} \angle M = f_\theta(X, D) \\
        & |\hat{Y}| = |M| \odot |O_B| \\
        & e^{i\angle \hat{Y}} = e^{i (\angle M + \angle O_B)} \\
        &\hat{y} = \mathcal{F}^{-1}(\hat{Y}) = \mathcal{F}^{-1}(|\hat{Y}|e^{i\angle \hat{Y}})
    \end{aligned}
\end{equation}
where $\mathcal{F}(\cdot)$ represents STFT, $f_\theta$ represents the neural network with learnable parameters $\theta$, and $\hat{Y}$ is the complex spectrogram of the estimated binaural signal $\hat{y}$.
\subsection{Model}
\label{sec:model}
We use a variety of neural network architectures to validate our framework, including Dense Neural Network (DNN), Gate Recurrent Unit (GRU) \cite{cho2014properties}, and UNet \cite{ronneberger2015u}. The DNN consists of 4 sequential fully-connected layers.
Each fully-connected layer is followed by a LeakyReLU with a negative sloop of 0.01. The output channels of all linear layers are: $1024$, $1024$, $128$, $6$ respectively. The final 6 channels output $|M|, \text{cos} \angle M, \text{sin} \angle M $ of the left and right ears.

GRUs are designed to be more suitable for handling sequential data
The used GRU consists of 4 layers: 3 bidirectional GRU layers and 1 linear layer. We flatten the channel and frequency dimension of the input feature to align with GRU layers. The output channel of all GRU layers is 1024 and the output channel of the linear layer is 6.

\begin{figure}[t]
  \centering
  \centerline{\includegraphics[width=0.7\columnwidth]{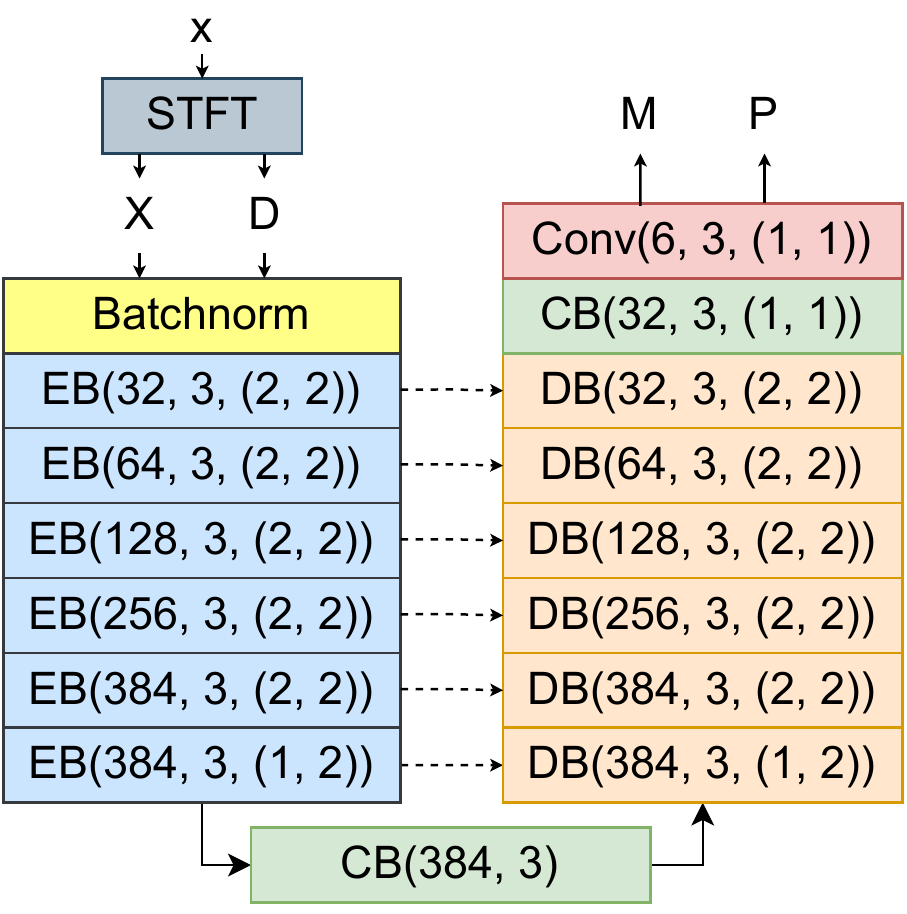}}
  \caption{The UNet architecture for binaural rendering.}
  \label{fig:model}
\end{figure}

CNNs are computationally more efficient than RNNs even with more layers, and the skip connection design of UNet \cite{ronneberger2015u} further improves training efficiency. The UNet consists of 6 encoder blocks and 6 decoder blocks as shown in Fig. \ref{fig:model}. The encoder block $EB(C_{\text{out}}, k, s)$ consists of a convolution block $CB(C_{\text{out}}, k)$ followed by an average pooling with a kernel size $s$. The convolution block $CB(C_{\text{out}}, k)$ consists of 2 convolution layers. Each convolution layer is followed by a batch normalization and a LeakyReLU, where $C_{\text{out}}$ is the number of output channels  and $k$ is the kernel size of both convolution layers. The decoder block $DB(C_{out}, k, s)$ consists of a transpose convolution layer with stride $s$ followed by a convolution block $CB(C_{\text{out}}, k)$. Finally, the feature passes a convolution block and an output convolution layer $Conv(C_{\text{out}}, k, s)$ where $s$ is the stride.
\subsection{Loss Function}
\label{sec:loss}
Experimentally, the time domain loss is easier to optimize. However, the pixel-to-pixel manner cannot portray the loss of phase differences. Therefore, we use an additional frequency domain loss as a compliment. Our loss function is the sum of L1 distances in both the time domain and the frequency domain:
\begin{equation}
\begin{aligned}
    L(\hat{y}, y) & = L_1^{\text{wav}}(\hat{y}, y) + \gamma L_1^{\text{sp}}(\hat{y}, y) \\
        & =||\hat{y} - y||_{1} + \gamma ||\mathcal{F}(\hat{y}), \mathcal{F}(y)||_{1},
\end{aligned}
\end{equation}
where $\gamma$ is a hyper-parameter adjusting the balance between the time domain loss and the frequency domain loss.

\section{Experiments}
\label{sec:pagestyle}

\subsection{Dataset and Experimental Settings}
We record and release a binaural rendering dataset publicly available\footnote{https://zenodo.org/record/7212795\#.Y1XwROxBw-R}. The dataset consists of paired ambisonic and binaural signals. Our recording devices include an H3-VR to record the first order of ambisonic and a Neumann KU100 dummy head to record the binaural signals. We arrange the devices in the center of a $20 m^2$ room and record 49 minutes of music played by a band. We split both ambisonic and binaural signals into 49 non-overlap and one-minute segments. There are 31 minutes of data for training and 18 minutes for evaluation. During training, we split those 1-minute segments into 3-second clips as the input to the binaural rendering models. We set $\gamma=1$ in loss function, use the Adam optimizer~ with $lr=0.001, \beta_1 = 0.9, \beta_2 = 0.999$, set the batch size to 16. The system trained for 100k steps, and took 13 hours on an Nvidia Tesla V100 32GB. 

\subsection{Object and Subjective Metrics}
We use three metrics to compare the performance of different systems: Signal-Distortion Ratio (SDR) \cite{lavrador2004evaluation}
defined in the time domain, Log Spectral Distance (LSD) \cite{enqvist2008minimal} defined in the frequency domain, and the Mean Opinion Score (MOS) for subjective evaluation. The SDR is defined as:
\begin{equation}
    SDR = 10 \log_{10} (\frac{y^2}{\hat{y}^2})
\end{equation}
where $y$ is the ground truth binaural signal and $\hat{y}$ is the estimated binaural signal. LSD is defined as:
\begin{equation}
    LSD = \sqrt{\frac{1}{2\pi}\int_{-\pi}^{\pi} [10 \log_{10} (\frac{Y(\omega)}{\hat{Y}(\omega)})]^2 d \omega}
\end{equation}
where $Y$ and $\hat{Y}$ are spectrograms of $y$ and $\hat{y}$ respectively, $\omega$ represents the frequency bin of the spectrogram. For subjective evaluation, MOS are scored on one test recording based on 4 dimensions: Q (quality), T (timbre), L (localization), and I (immersion). The MOS ranges from 1 to 5 where 5 is the best. Our main reference is based on the average value of MOS scores. The standard deviation is shown in parentheses.

\subsection{Overall Comparison}
Table \ref{table:EXP1} compares the sp-HRTF-based method and our proposed nn-based methods. The sp-HRTF system is optimized by MagLS \cite{schorkhuber2018binaural} with the SADIE dataset \cite{app8112029}. Table \ref{table:EXP1} shows that all NN-based methods are better than the sp-HRTF method in both SDR and LSD. Second, the sp-HRTF method performs better than the DNN in all MOS dimensions. This is mainly due to the insufficient capacity of the DNN. Third, both GRU and UNet perform better than the sp-HRTF method in sound localization. The GRU system achieves the highest MOS scores Q, T, L, and I of 3.83, 3.58, 3.87, 3.58, respectively. One reason is that the recurrent design of GRU fits binaural rendering. Last, we found that a proper data pre-processing improves SDR and LSR significantly. The sixth row of Table \ref{table:EXP1} mixs other 2 clips for each clip and increases SDR to 9.17 from 7.32. The seventh row mixs other 4 clips for each clip and further increases SDR to 9.86. However, the improvement of SDR is at the cost of MOS scores.
\begin{table}[ht]
\centering
\caption{Evaluation metrics of different systems}
\resizebox{\columnwidth}{!}{%
\begin{tabular}{*{7}{c}}
 \toprule   
 & SDR & LSD & Q & T & L & I\\
 \midrule
Oracle & $\infty$ & 0 & 4.5 (0) & 4.50 (0) & 4.5 (0) & 4.5 (0) \\
sp-HRTF & -0.79 & 1.88 & 3.58 (0.45) & 2.67 (0.75) & 3.37 (0.72) & 3.25 (0.69) \\
 \midrule
 DNN-4 & 5.58 & 0.95 & 1.78 (0.66) & 2.25 (0.99) & 3.08 (1.01) & 2.17 (0.90) \\
 GRU-4 & 7.32 & 0.95 & \textbf{3.83} (0.37) & \textbf{3.58} (0.73) & \textbf{3.87} (0.39) & \textbf{3.58} (0.45) \\ 
 UNet-41 & 8.03 & 0.93 & 3.5 (0.65) & 3.47 (0.47) & 3.95 (0.21) & 3.5 (0.29)\\
 GRU-4 (2 mix) & 9.17 & 0.88 & 3.42 (0.19) & 3.33 (0.47) & 3.42 (0.45) & 3.33 (0.37) \\
 GRU-4 (4 mix) & \textbf{9.86} & \textbf{0.85} & 3.42 (0.45) & 3.33 (0.47) & 3.42 (0.53) & 3.20 (0.58) \\ 
 \bottomrule
\end{tabular}}
\label{table:EXP1}
\end{table}
\subsection{Evaluation of Input Features}
Table \ref{table:EXP2} compares GRU systems trained with different input features. On one hand, the first row uses the magnitude spectrogram as the input feature and achieves the best performance on SDR and LSD. On the other hand, our proposed input feature outperforms the magnitude spectrogram input feature in all MOS dimensions, especially in sound localization and immersion feeling. This indicates that phase differences are important in perceptual evaluation.
\begin{table}[ht]
\centering
\caption{Evaluation metrics of input features}
\resizebox{\columnwidth}{!}{%
\begin{tabular}{*{7}{c}}
 \toprule
 & SDR & LSD & Q & T & L & I\\
 \midrule
Mag. & \textbf{7.45} & \textbf{0.91} & 3.45 (0.31) & 3.50 (0.29) & 3.42 (0.34) & 3.00 (0.50)\\    
Complex & 6.69 & 0.99 & 2.97 (0.30) & 3.08 (0.61) & 2.5 (0.82) & 2.83 (0.62)\\
Complex + Rel. Phase * & 7.32 & 0.95 & \textbf{3.83} (0.37) & \textbf{3.58} (0.73) & \textbf{3.87} (0.39) & \textbf{3.58} (0.45) \\
 \bottomrule
\end{tabular}}
\label{table:EXP2}
\end{table}
\subsection{Evaluation of Output Features}
Table \ref{table:output} compares different GRU systems trained with different output features. The first row outputs the complex spectrum directly and achieves MOS of around 1.1 in all Q, T, L, and I. The second row decouples the magnitude and phase and achieves Q, T, I of around 2.1 and L of 3.17. 
\begin{table}[ht]
\centering
\caption{Evaluation metrics of output features}
\resizebox{\columnwidth}{!}{%
\begin{tabular}{*{7}{c}}
 \toprule
 & SDR & LSD & Q & T & L & I\\
 \midrule 
Complex & 4.61 & 1.15 & 1.08 (0.19) & 1.17 (0.37) & 1.08 (0.19) & 1.08 (0.19)\\
Mag. + Rel. phase & 7.28 & 1.07 & 2.08 (0.53) & 2.08 (0.34) & 3.17 (0.69) & 2.17 (0.62) \\
Mask + Rel. phase * & \text{7.32} & \textbf{0.95} & \textbf{3.83} (0.37) & \textbf{3.58} (0.73) & \textbf{3.87} (0.39) & \textbf{3.58} (0.45) \\ 
 \bottomrule
\end{tabular}}
\label{table:output}
\end{table}
\subsection{Evaluation of loss functions}
Table \ref{table:loss} compares different GRU systems trained with different loss functions. The first row using only the time domain loss raises SDR from 7.32 to 7.35 and increases LSD from 0.95 to 1.29. The second row using only the frequency loss decreases SDR from 7.32 to -2.31 and decreases LSD from 0.95 to 0.94. This shows that optimizing the time domain loss is easier than the frequency domain loss. 
The third row shows that the L1 distance is better than the L2 distance in all metrics. 
\begin{table}[ht]
\centering
\caption{Evaluation Metrics of Different Loss Functions}
\resizebox{\columnwidth}{!}{%
\begin{tabular}{*{7}{c}}
 \toprule
 & SDR & LSD & Q & T & L & I\\
 \midrule 
    $L_1^{wav}$ & \textbf{7.35} & 1.29 & 2.42 (0.61) & 2.75 (0.63) & 2.58 (0.98) & 2.58 (0.61) \\
    $L_1^{sp}$ & -2.31 & \textbf{0.94} & 2.97 (0.70) & 3.33 (0.47) & 2.83 (0.80) & 2.67 (0.69)\\
    $L_2^{wav} + L_2^{sp}$ & 4.83 & 1.08 & 2.83 (0.62) & 3.00 (0.76) & 3.17 (0.69) & 2.58 (0.45)\\  
    $L_1^{wav} + L_1^{sp}$ * & 7.32 & 0.95 & \textbf{3.83} (0.37) & \textbf{3.58} (0.73) & \textbf{3.87} (0.39) & \textbf{3.58} (0.45) \\ 
 \bottomrule
\end{tabular}}
\label{table:loss}
\end{table}
\section{Conclusion}
We collect release an ambisonic-binaural dataset, and propose a deep learning framework for binaural rendering from ambisonic signals. 
We show that all DNN, GRU, and UNet networks outperform conventional methods in both SDR and LSD. The GRU even achieves the best MOS socres on all dimensions at the cost of slight loss of SDR and LSD. We show that mixing data augmentation is helpful for improving SDR and LSD, but may decrease MOS scores. We evaluate different input features, output features, and loss functions for neural network-based binaural rendering. 
In future, we will focus on designing new architectures for NN-based binaural rendering.

\section{Acknowledgement}
We are grateful to Billowing Fairy Tale band for their help in recording the data set, and to Menglong Feng and Yuhao Wang for their help in scoring MOS.

\bibliographystyle{IEEEbib}
\bibliography{refs, related}

\end{document}